\documentclass{PoS}
\title{Fisher's zeros, complex RG flows and confinement in LGT models.}

\ShortTitle{Fisher's zeros, complex RG flows and confinement in LGT models.}

\author{\speaker{Alan Denbleyker}$;^a$\footnote{Current email address: alan-denbleyker@gmail.com}, Alexei Bazavov $^b$,
Daping Du $^{a,c,d}$, Yuzhi Liu $^a$, Yannick Meurice $^a$, Haiyuan Zou $^a$\\
\llap{$^a$} Department of Physics and Astronomy, University of Iowa, Iowa City, IA 52240, USA\\
\llap{$^b$} Brookhaven National Laboratory, Upton, NY 11973, USA\\
\llap{$^c$} Fermi National Accelerator Laboratory, Batavia, IL 60510, USA \\
\llap{$^d$} Physics Department, University of Illinois, Urbana, IL 61801, USA\\
}

\abstract{The zeros of the partition function in the complex $\beta$ plane (Fisher's zeros) play an important role in our understanding of phase transitions and RG flows. Recently, we argued that they act as gates or separatrices for complex RG flows. Using histogram reweighting to construct the density of states, we calculate the Fisher's zeros for pure gauge $SU(2)$ and $U(1)$ on $L^4$ lattices. For $SU(2)$, these zeros appear to move almost horizontally when the volume increases. They stay away from the real axis which indicates a confining theory at zero temperature. We discuss the effect of an adjoint term on these results. In contrast, using recent multicanonical simulations for the $U(1)$ model for $L$ up to 8 we find that the zeros pinch the real axis near $\beta$=1.0113. Preliminary results concerning $U(1)$ at larger volumes, $SU(3)$ with 3 light flavors and plans to delimit the boundary of the conformal window are briefly discussed.
}

\FullConference{XXIX International Symposium on Lattice Field Theory \\
                July 10-16, 2011\\
                Squaw Valley, Lake Tahoe, California}

\begin{document}

\section{Introduction}
Given the successes of the local gauge invariance principle and recent experimental results, the possibility that new gauge interactions are responsible for electro-weak symmetry breaking looks quite attractive. One particularly interesting situation from a phenomelogical point of view \cite{Sannino:2009za,DeGrand:2010ba} is when the Callan-Symanzik $\beta$ function for the new gauge coupling  approaches zero from below and the running coupling constant encounters only small changes over a significant range of scale.  We then say that the ``running'' coupling constant ``walks''. 

This nearly conformal situation can be reached by tuning a parameter (typically the number of light fermions)  in such a way that the zeros of the $\beta$ function (and the corresponding  fixed points of the Renormalization Group (RG) transformation) disappear in the complex plane. Other models where conformality can be lost and restored by tuning a parameter (for instance the quantum mechanical $1/r^2$ potential) have been studied recently \cite{Kaplan:2009kr,moroz09}.  

This motivated us to study extensions of the RG flows in the complex coupling plane \cite{Letter10,PhysRevD.83.056009,Liu:2011zzh}.  A general feature that we observed is that the Fisher's zeros - the zeros of the partition function in the complex $\beta$ plane - act as  ``gates'' for the RG flows ending at the strongly coupled fixed point. This can be seen as a complex extension of the general picture proposed by Tomboulis \cite{Ogilvie:2010vx,Tomboulis:2009zz}: in confining theories, the gate stays open as the volume increases and  RG flows starting in a complex neighborhood the UV fixed point  where asymptotic freedom is working, can reach the IR fixed point where confinement and the existence of a mass gap are clearly present. Furthermore, we observed \cite{Liu:2011zzh} that 
the discrete RG transformation approximately maps Fisher's zeros for a given lattice size into the zeros of for the blocked lattice size, forming separatrices among the flows ending at different IR fixed points. In short, the global properties of the RG flows can be determined by calculating Fisher's zeros, bypassing explicit calculations of the RG flows which are technically difficult and have lattice artifacts that are sometimes difficult to decipher.  These results are reviewed in section \ref{sec:spin}. 

Recently, we have developed new methods to calculate the Fisher's zeros in lattice gauge theory without fermions \cite{Letter10,quasig,Bazavov:2009wz,Bazavov:2010xh,su2progress,abprogress}. The methods rely on the construction of the density of states \cite{Denbleyker:2008ss} and its analytical continuation to the complex energy plane. These methods have been applied to the case of $U(1)$ and $SU(2)$ and provide a clear picture of the large scale behavior of these models. These results are summarized in section \ref{sec:gauge}. 
As briefly explained in section \ref{sec:prel}, we started to pursue this effort in the case of $SU(3)$ with various number of fermions and plan to provide new criterions to delimit boundary of the conformal window. 

\section{Complex RG Flows in Spin Models}\label{sec:spin}

Calculations of RG flows and discrete $\beta$ functions in lattice gauge theory are notoriously difficult. One of the main question asked is  how many flavors does it take to destroy the confining properties of the theory. For gauge theories, the absence of long-range order (no massless gluons) is associated with confinement. The absence of long-range order also characterizes the 2-dimensional $O(N)$ sigma models with $N\leq 3$. By using some approximations, it is possible to calculate complex RG flows and Fisher's zeros much more easily than for gauge theories. 

Recently \cite{Liu:2011zzh}, we were able to illustrate how RG fixed points can disappear in the complex inverse temperature ($\beta$)  plane using the two-lattice matching procedure \cite{PhysRevB.27.1736,Hasenfratz:1984hx} for Dyson's  hierarchical model \cite{dyson69,baker72} with an Ising measure. In this model, the local potential approximation is exact and RG flows can be calculated numerically with good accuracy \cite{hmreview}. 

The model has a free parameter that plays the role of the dimension and can be tuned continuously. For $D=3$, the model has a nontrivial Wilson-Fisher fixed point. As we lower $D$ continuously the fixed point on the real axis moves to the right and disappears at infinity for $D=2$ in agreement with a rigorous results \cite{hmreview}. 
  
\begin{figure}
  \includegraphics[angle=-90,width=3.in]{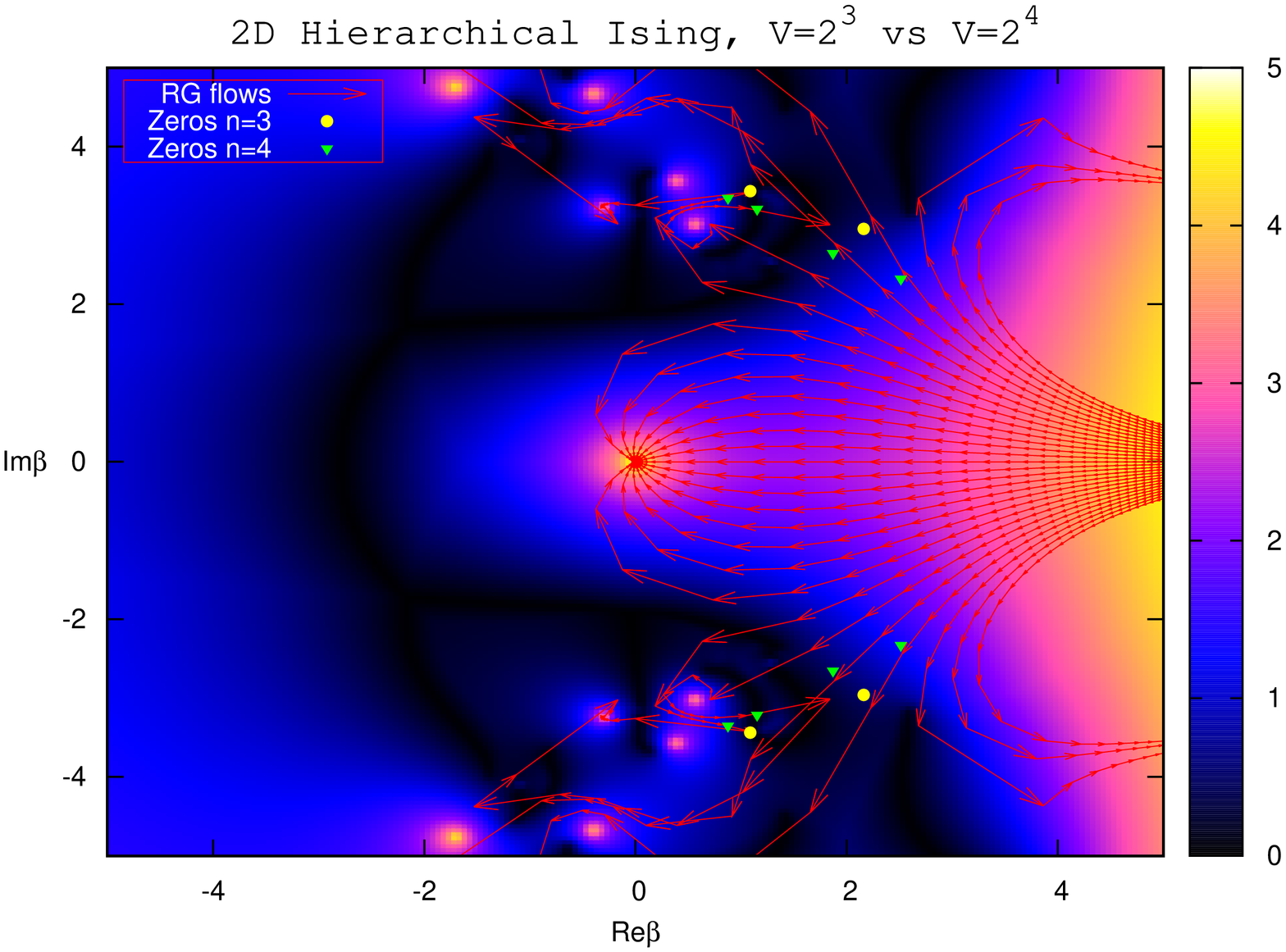}
  \includegraphics[angle=-90, width=3.in]{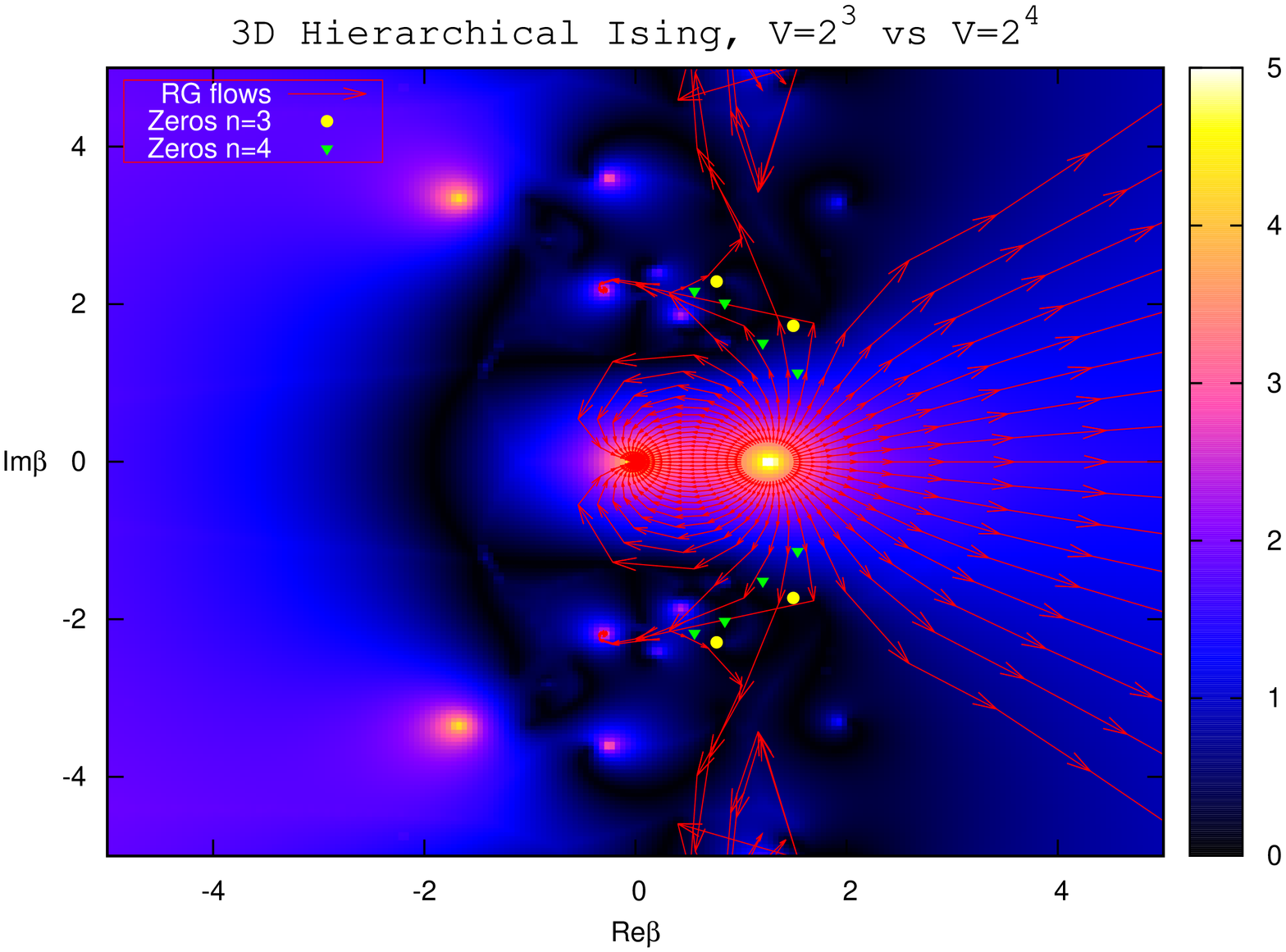}
  \caption{\label{fig:chm} Complex flows for the hierarchical Ising model for $D=2$ (left) and $D=3$ (right). See \cite{Liu:2011zzh} for details.}
\end{figure}

The complex RG flows are shown in Fig. \ref{fig:chm}. On the left, for $D=2$, the complex flow lines go smoothly from infinity to zero, which indicates that the system has no phase transition. For $D=3$, the  complex flow line start from $\beta_c$ and end to either zero or infinity. The darker region of the graph signal competing solutions for the matching condition (see  \cite{Liu:2011zzh} for details). 

\begin{figure}
\begin{center}
  \includegraphics[width=4in]{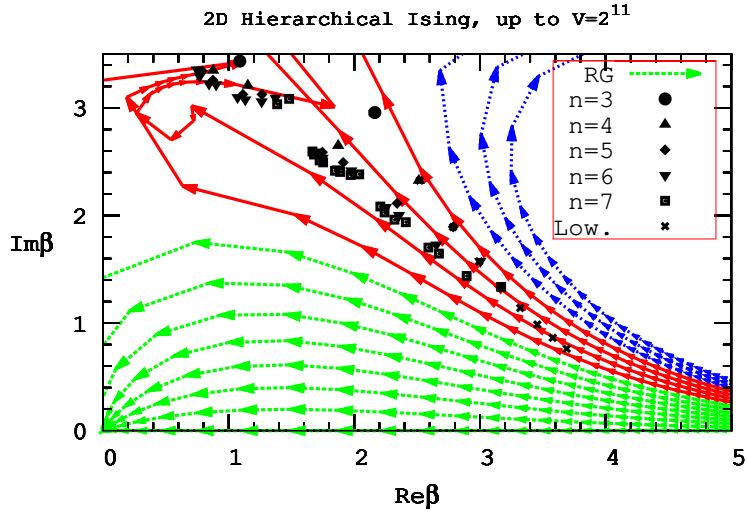}
  \caption{\label{fig:small} $D=2$ Fisher's zeros for larger numbers of sites. See \cite{Liu:2011zzh} for details.}
  \end{center}
\end{figure}

An important feature that can be observed on this figure is that  the RG flow approximately follow the Fisher zeros along the separatrice between different basins of attraction. More specifically, in one discrete step, the RG flows approximately go from the the zeros at one given number of lattice sites to the zeros for a smaller number of sites as obtained after block spinning the first lattice. This approximate property is illustrated for Fisher's zeros of  higher volumes in Fig. \ref{fig:small}. This property was found to be exact for the 1-dimensional Ising model at complex temperature  \cite{Damgaard:1993df} for which decimation can be performed exactly. In the case of our calculation, the reduction to one-dimensional flows used for the matching is only approximate since a finite number of block spin only partially eliminates the irrelevant directions. 

This example shows that the global properties of the RG flows (difficult to calculate) can be inferred from the location of the Fisher's zeros at successive volume (more easy to calculate). Similar observations were made for the $2D\ O(N)$ non-linear sigma models in the large-$N$ limit \cite{Letter10,PhysRevD.83.056009}. We constructed the Riemann sheet structure and singular points of the finite lattice size $L$ mappings between the mass   gap and the 't Hooft coupling. We argued that the Fisher's zeros appear on  ``strings" ending approximately at the singular points. We compared finite  volume complex flows obtained from the rescaling of the ultraviolet cutoff in the gap equation and from the two lattice matching. In both cases, the flows are channelled through the singular points and end at the strong coupling fixed points, however strong scheme dependence appear on the ultraviolet side.

\section{Fisher's Zeros in Lattice Gauge Theory}\label{sec:gauge}

We used the density of states  $n(S)$ to study the zeros of pure $SU(2)$ and pure $U(1)$ 4D gauge theories. The partition function can be written as 
\begin{equation}
  Z(\beta) = \int dS n(S) e^{-\beta S} \  .
\end{equation} 
The corresponding entropy density function is $f(x) = \ln n(\mathcal{N}_p x)/ \mathcal{N}_p$, where $\mathcal{N}_p\equiv 6\times L^4$ is the total number of plaquettes. As the volume increases, the complex partition function zeros will in some cases pinch the real axis  at the transition point. For a second order phase transition, the imaginary part of the lowest zero decreases with a power related to the critical exponent $\nu$ :
 \begin{equation}
  Im\beta (L) \sim L^{-1/\nu}\ .
\end{equation}
The formula also holds for a first order phase  transition, provided that we replace $\nu$ by $1/D$. 

\begin{figure}
\begin{center}
  \includegraphics[width=2.8in]{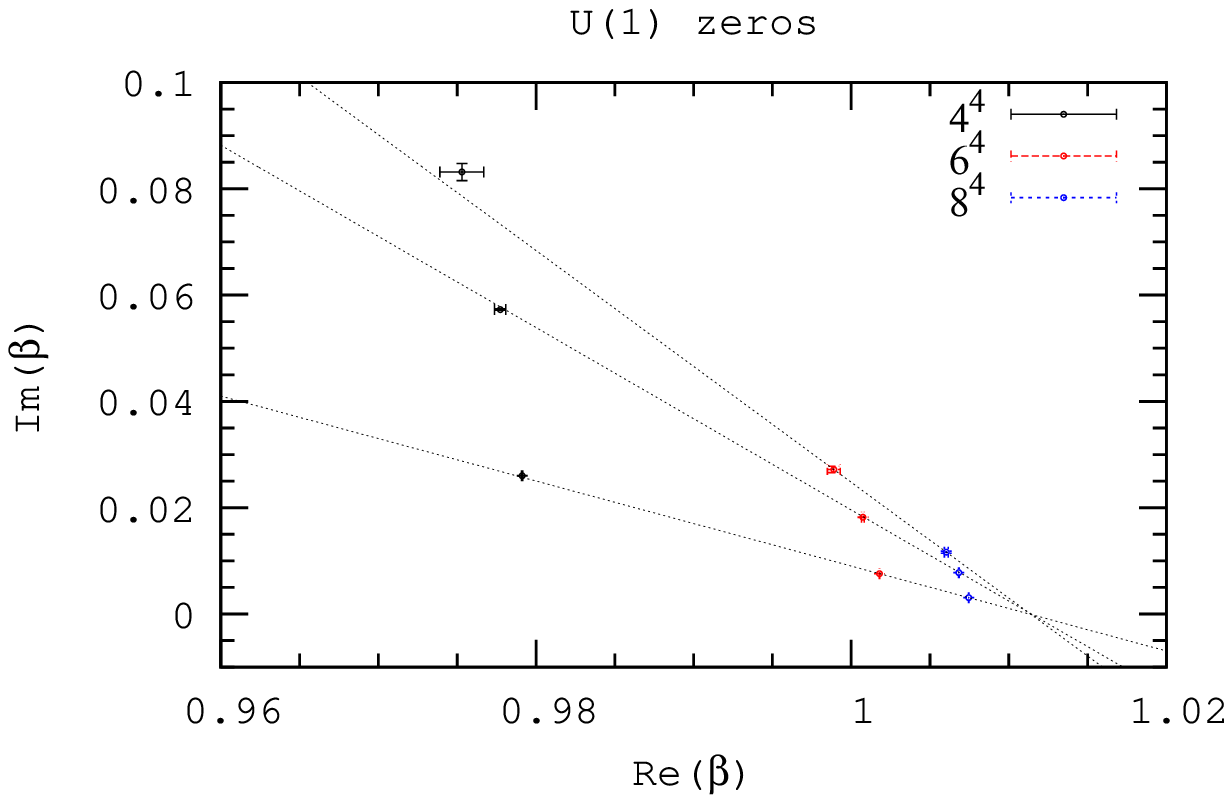}
   \includegraphics[width=2.8in]{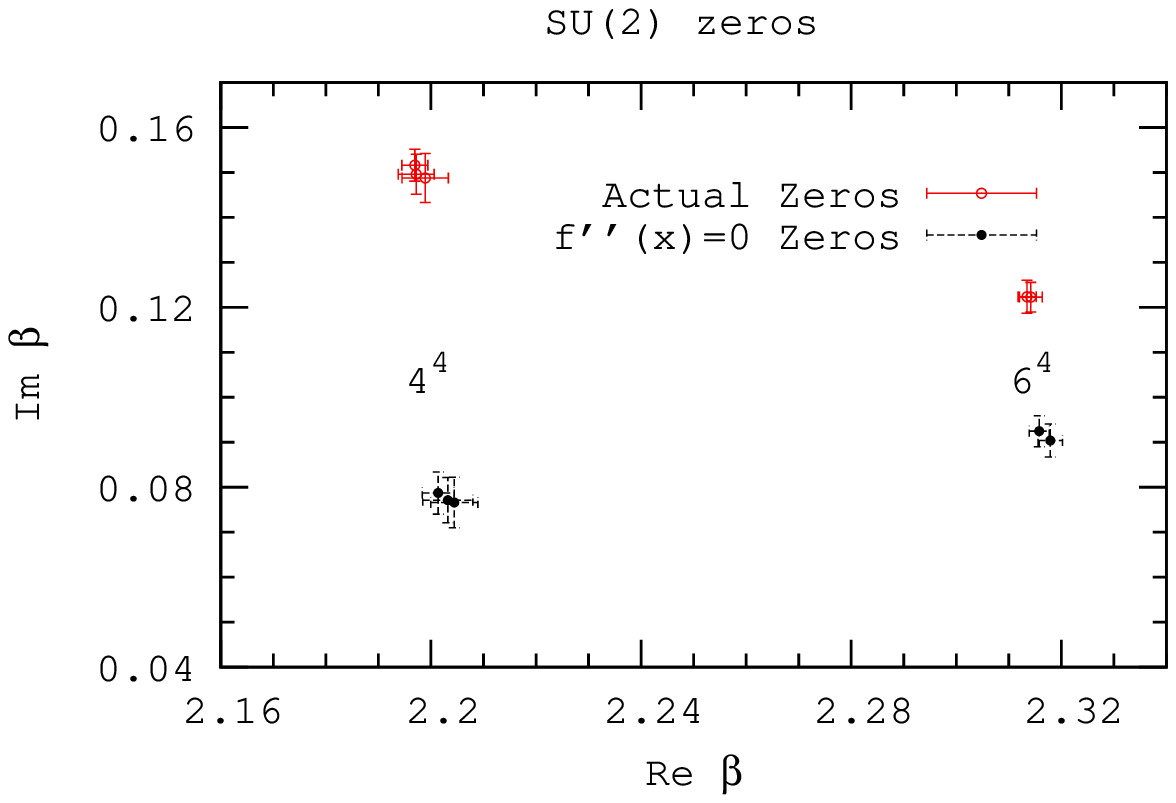}
  \caption{Fisher's zeros for $U(1)$ (left) and $SU(2)$ (right) .}
  \end{center}
  \label{U1graphs}
\end{figure}

In the case case of a pure  $U(1)$ gauge theory, we used the multi-canonical  algorithm for lattices with sizes $L=4,6,8$ \cite{Bazavov:2009wz,Bazavov:2010xh} to construct the density of states. We calculated the lowest three zeros for these volumes (see Fig. \ref{U1graphs}).  We were able to locate the lowest zeros with a precision of order $10^{-5}$.  The zeros cross at $\beta_c=1.0113(2)$. The imaginary part of the zeros scales with $L^{-3.07}$,  or $\nu = 0.326$, possibly consistent with a second order phase transition. However the zeros from higher volumes show that the scaling is "rolling" and $\nu$ is  decreasing with  the volume. This will be discussed in a forthcoming preprint \cite{abprogress}. 

\begin{figure}
  \includegraphics[width=3in]{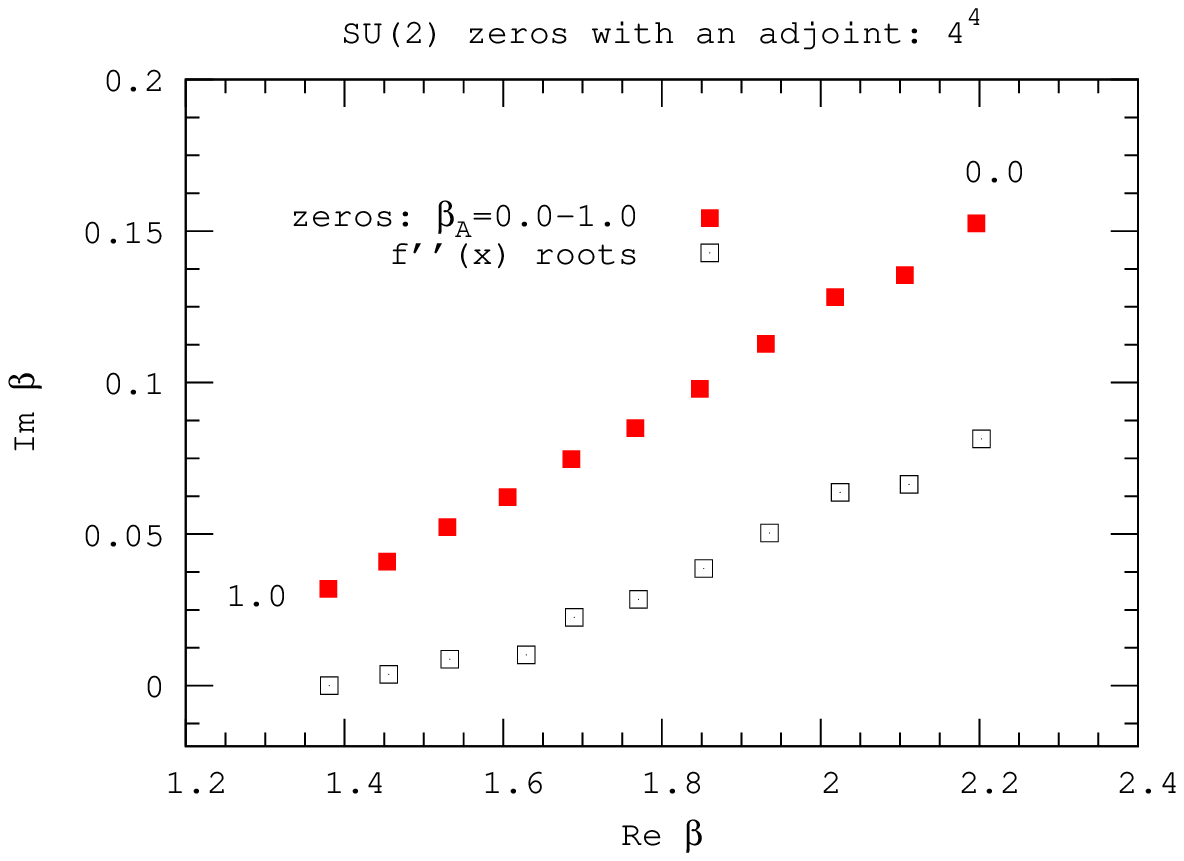}
  \includegraphics[width=3in]{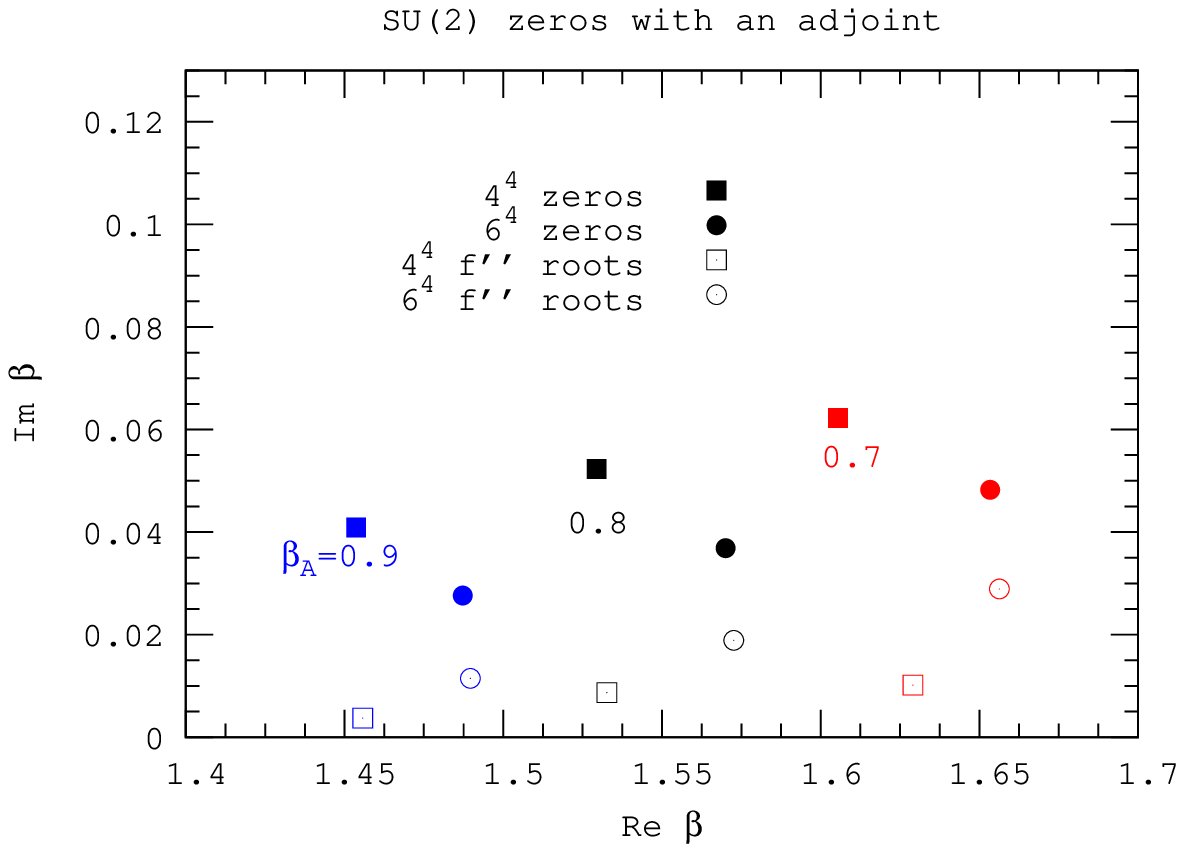}
  \caption{Fisher's zeros in SU(2) with an adjoint term for different volumes, and compared to roots of $f^{\prime\prime}$.}
  \label{fig:z4}
\end{figure}

For pure  gauge $SU(2)$ with a Wilson action,  the lowest complex zeros for  the volumes $4^4$ and $6^4$ are shown in Fig. \ref{fig:z4}. The upper points (red) are the locations of the partition function zeros, while the lower points are the complex roots of $f''(x)$ mapped to the $\beta$-plane using $f'$. The multiple points at each location correspond to three different ranges of the Chebyshev parametrization of $f(x)$. The error bars reflect the statistical uncertainty and were obtained by comparing the results based on independent simulations. 

When an adjoint term is added for SU(2), the action reads
\begin{equation}
  S_\square = \beta (1 - \frac{1}{2}Tr U_\square) + \beta_A \left[1 - 
  \frac{1}{3}(|Tr U_\square )|^2-1) \right] \  .
\end{equation} 
The top right graph in Fig. \ref{fig:z4} shows the complex zeros (solid squares) at volume $4^4$ with $\beta_A=0.0$ to $1.0$ with an increment $0.1$. The roots of $f''(x)$ (empty squares) with $f(x)$ calculated at a finite volume $L^4$ correspond to the zeros as if $L\to\infty$ was taken.  The actual zeros approach the roots of $f''(x)$ as the volume increases and coincide with them in the $L\to \infty$ limit. The right graph in Fig. \ref{fig:z4} shows the zeros and the roots of $f''$ in the volume $4^4$ and $6^4$ with $\beta_A=0.7,0.8$ and $0.9$. The rising of the $f''(x)$ roots with volume indicates the SU(2) zeros will stabilize at a distance away from the real axis. 

\section{Conclusions and perspectives}\label{sec:prel}

\begin{figure}[h]
\begin{center}
  \includegraphics[width=4in]{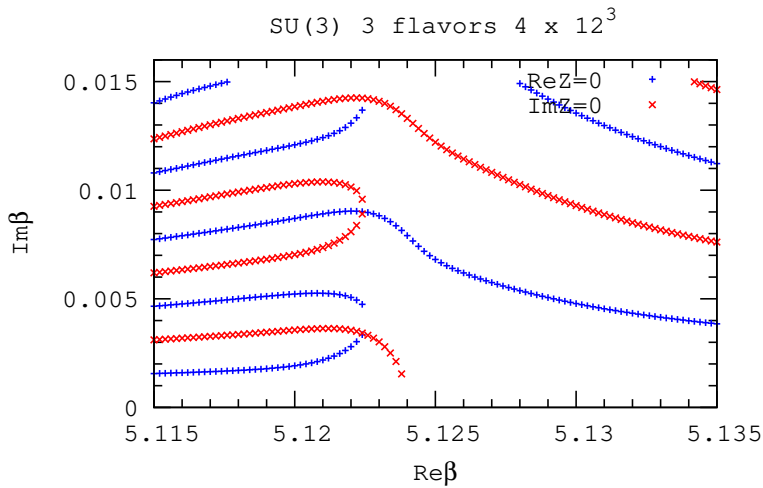}
  \caption{\label{fig:33}Curves for the  zeros of the real and imaginary part of the partition function in the complex plane for $SU(3)$ with three light flavors on a $4\times 12^3$ lattice.}
  \end{center}
\end{figure}

In conclusions, we have shown that the stabilization of Fisher's zeros away from the real axis can be used as a signature for 
a confining theory. We plan to apply this method to locate the boundary of the conformal window in multiflavor models. 
As a first step we studied the case of $SU(3)$ with 3 light quarks at finite temperature and found a scaling $L^{-3}$ 
 for the imaginary part consistent with a first order phase transition. The calculations were done with unimproved 
  staggered fermions with $m=0.02$. The zeros are shown in Fig. \ref{fig:33}. 

\section{Acknowledgment}

We thank D. K. Sinclair for providing high statistics $SU(3)$ plaquette data for analysis. We thank M. B. Oktay for helping with MILC fermion codes.

\end{document}